\documentclass[floatfix,prb,twocolumn]{revtex4}
\usepackage{graphicx}

\usepackage{tikz}
\usepackage{ifthen}
\usepackage{xcolor}
\definecolor{DarkGreen}{rgb}{0,0.7,0}
\definecolor{DarkRed}{rgb}{0.5,0.0}

\newcommand{\ColorOnline}{(Color on-line) }

\newcommand{\HH}{\ensuremath{{\cal H} }}
\newcommand{\HLead}{\ensuremath{{\cal H}_{\mathrm{lead}}}}
\newcommand{\HRing}{\ensuremath{{\cal H}_{\mathrm{Ring}}}}
\newcommand{\HContact}{\ensuremath{{\cal H}_{\mathrm{Contact}}}}

\newcommand{\tDot}{\ensuremath{{t}_{\mathrm{Dot}}}}
\newcommand{\tL}{\ensuremath{{t}_{\mathrm{L}}}}
\newcommand{\tR}{\ensuremath{{t}_{\mathrm{R}}}}
\newcommand{\tBC}{\ensuremath{{t}_{\mathrm{BC}}}}
\newcommand{\MS}{\ensuremath{{M}_{\mathrm{S}}}}
\newcommand{\TK}{\ensuremath{{T}_{\mathrm{K}}}}

\newcommand{\e}{{\mathrm{e}}}
\newcommand{\im}{{\mathrm{i}}}
\newcommand{\Vg}{V_{\mathrm{gate}}}

\graphicspath{{.}} 

\begin{document}


\title{The dark side of benzene: interference vs.\ interaction}

\author{Dan Bohr}
 \affiliation{Institut f\"ur Theorie der Kondensierten Materie, Karlsruher Institut f\"ur Technologie, 76131 Karlsruhe, Germany}%
\author{Peter Schmitteckert}%
 \affiliation{Institut f\"ur Nanotechnologie, Karlsruher Institut f\"ur Technologie, 76344 Eggenstein-Leopoldshafen, Germany}%
 \affiliation{DFG Center for Functional Nanostructures, Karlsruhe Institute of Technology, 76128 Karlsruhe, Germany}

\date{\today}

\begin{abstract} 
We present the study of the linear conductance vs.\ applied gate voltage for an interacting six site ring structure,
which is threaded by a flux of $\pi$ and coupled to a left and a right lead.
This ring structure is designed to have a  vanishing conductance for all gate voltages and temperatures
provided interactions are ignored. Therefore this system is an ideal testbed to study the interplay of interaction and interference.
First we find a Kondo type resonance for rather large hopping parameter. Second, we find additional resonance peaks which 
can be explained by a population blocking mechanism.
To this end we have to extend the Kubo approach within the Density Matrix Renormalization Group method to handle degenerate states.
\end{abstract}

\maketitle %

\section{Introduction}
The continuing miniaturization trend in solid state physics has lead to the discovery of significant new effects 
and interesting properties of new devices. Two effects are playing major roles: Interaction and interference. 
Interference is the essential component in  Aharonov-Bohm rings, whereas interaction is the essential 
component in Coulomb blockade experiments and Kondo physics. Combining these two effects has turned out to be a formidable 
theoretical task.\cite{Begemann_Darau_Donarini_Grifoni:2008,Wingreen_et_al_1993,Meir_et_al_1993,Pedersen_Bohr_Wacker_Novotny_Schmitteckert_Flensberg:2009} 
The appearance of a Fano anti resonance has been suggested as a possibility for an interference based transistor \cite{Stafford_et_al:2007}
and discussed in more detail in \cite{Ke_Yang:2008}.
The role of interference effects for transport properties even for acyclic molecules has been discussed in \cite{Salomon:2008}.
Going beyond perturbative approaches the interplay of interaction and interference in the transport properties  
was studied by Roura-Bas et. al \cite{Aligia_Hallberg:2011}.
There the authors found a crossover between a SU(2) and SU(4)Anderson model mediated by an external level splitting.

In this work we employ the density matrix renormalization group (DMRG) method\cite{White:1992,White:1993} to investigate an example of a structure 
where both interference and interaction play a strong role, a spinless tight-binding model of a ring-structure. 
Threading half a quantum of flux through the ring, and disregarding interactions the conductance of this specific ring is analytically zero for all gate voltages
as each single-particle level is two-fold degenerate and displays perfect destructive interference, see appendix.

Including interaction this picture changes qualitatively: A very broad resonance appears as the strength of the interaction is increased. 
Introducing an asymmetry in the ring gradually destroys this resonance, and a ``normal'' Lorentzian line shape at a 
Coulomb blockade position appears,  where the single particle levels  of the isolated ring are half occupied.

An exponentially decaying line shape combined with a conductance plateau indicates that the conductance peak at small gate voltages
is not described by a simple single particle picture and is similar to conductance plateaus in Kondo physics. 
Breaking the degeneracy by perturbing a single hopping-matrix element on the ring destroys this conductance peak.
In contrast to usual Kondo physics new conductance peaks appear which we explain by a level/interference blocking mechanism.
In contrast to the model studied in \cite{Aligia_Hallberg:2011} the Kondo effect discussed in work is stabilized by the charge
density wave like ordering due to the umklapp scattering in the presence of the underlying lattice.

\section{Model}
\begin{figure}[htb]
\begin{tikzpicture}

	\newdimen\a
	\a=0.75cm
	\newdimen\b
	\b=0.5\a
	\newdimen\c
	\c=0.86603\a
	\newdimen\d
	\d=0\a
	\newdimen\dia
	\dia=0.15\a

	\draw (0,0) -- ++(\b,\c) -- ++(\a,\d) -- ++(\b,-\c) -- ++(-\b, -\c) -- ++(-\a,\d) -- ++(-\b,\c);
	\draw[color=red] (0,0) -- ++(\b,\c);

	\foreach \x in {0,...,2} {
	\draw (-\x *\a,0) -- ++(-\a,0);
	};
	\foreach \x in {0.0, 0.2214, 0.4918, 0.8221, 1.2255} {
	\draw (-3*\a, 0) -- (-4*\a, \x *2*\a) ;
	\draw (-3*\a, 0) -- (-4*\a, -\x *2*\a) ;
	\fill[color=DarkGreen] (-4*\a, \x *2*\a) circle (\dia);
	\fill[color=DarkGreen] (-4*\a, -\x *2*\a) circle (\dia);
	};
	\foreach \x in {0,...,2} {
	\fill[color=DarkGreen] (-\x *\a -\a,0) circle (\dia);
	};

	\foreach \x in {0,...,2} {
	\draw (\a +2*\b +\x *\a,0) -- ++(\a,0);
	};
	\foreach \x in {0.0, 0.2214, 0.4918, 0.8221, 1.2255} {
	\draw (5*\a, 0) -- (6*\a, \x *2*\a) ;
	\draw (5*\a, 0) -- (6*\a, -\x *2*\a) ;
	\fill[color=DarkGreen] (6*\a, \x *2*\a) circle (\dia);
	\fill[color=DarkGreen] (6*\a, -\x *2*\a) circle (\dia);
	};
	\foreach \x in {0,...,2} {
	\fill[color=DarkGreen] (\a +2*\b +\x *\a +\a,0) circle (\dia);
	};

	\fill[color=blue] (0,0) circle (\dia); 
	\fill[color=blue] (\b,\c) circle (\dia);
	\fill[color=blue] (\a+\b,\c) circle (\dia);
	\fill[color=blue] (\a+2*\b,0) circle (\dia); 
	\fill[color=blue] (\a+\b,-\c) circle (\dia);
	\fill[color=blue] (\b,-\c) circle (\dia);
\end{tikzpicture}
\caption{\ColorOnline Illustration of the finite size setup used in the DMRG calculations. 
The sites in the ring are shown in blue, and the implementation of the flux, $\tBC$, 
is shown as a red hopping between sites $1$ and $6$ on the ring. 
Each lead (shown in green) is described by a real-space part coupled to a momentum-space part. 
The real-space part of the leads ensures a proper representation of local physics, 
whereas the momentum-space part ensures that the low-energy spectrum is properly represented. 
Note that the outermost real-space site in each lead is coupled to \emph{all} momentum-space sites of that lead.}\label{Ring-Lead-Setup}
\end{figure}
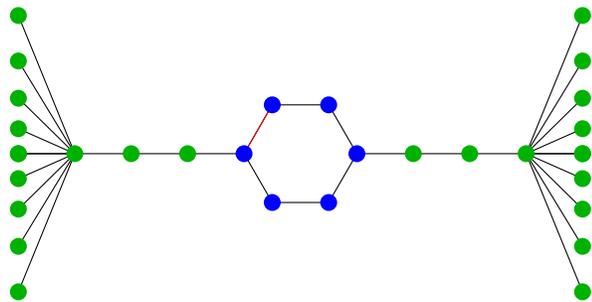

In this paper we consider a model consisting of a (hexagonal) ring of sites, through which is threaded half a flux quantum. 
The ring is then coupled symmetrically to leads in a two-terminal setup, as indicated in Fig.~\ref{Ring-Lead-Setup}. 
The flux through the ring is modeled using Peierls substitution as a phase on the hopping-matrix elements on the ring, 
such that the relative phase of the hopping elements on the ring changes by $\pi$ going once around the ring. 
The distribution of this phase over the ring is insignificant as it can always be gauged into a single bond. 
In this work we choose to modify a single hopping element only, 
denoted $\tBC$, as indicated in red in Fig.~\ref{Ring-Lead-Setup}. Including a particle-hole symmetric nearest-neighbor density-density 
interaction on the ring the Hamiltonian of the model is
\begin{equation}
	\HH = \HRing + \HLead + \HContact,
\end{equation}
\begin{widetext}
\begin{eqnarray}
	\HRing  &=& \sum_{i=1}^{\MS-1} \left (t^{}_{i} d_i^\dag d^{}_{i+1} + t_{i}^* d_{i+1}^\dag d^{}_{i} \right) + \tBC^{} d_{\MS}^\dag d^{}_1 + \tBC^* d_{1}^\dag d^{}_{\MS}\nonumber \\
	&& + \sum_{i=1}^{\MS-1} U_i \Big(n_i-\frac 12\Big) \Big(n_{i+1}-\frac 12\Big) + U_{\MS} \Big(n_{\MS} -\frac 12\Big) \Big(n_1-\frac 12\Big) + \sum_{i=1}^{\MS} V_{\mathrm{gate}} n_i \\
	\HLead    &=& \sum_{i=1}^{M}\sum_{\alpha=L,R} \left(t^{}_{i,\alpha} c_{i,\alpha}^\dag c^{}_{i+1,\alpha} + t_{i,\alpha}^* c_{i+1,\alpha}^\dag c^{}_{i,\alpha} \right) + \sum_k\sum_{\alpha=L,R} \Big(\epsilon^{}_{k,\alpha} n^{}_{k,\alpha}  + t^{}_{k,\alpha} c_{M,\alpha}^\dag c^{}_{k,\alpha} + t_{k,\alpha}^* c_{k,\alpha}^\dag c^{}_{M,\alpha} \Big) \\
	\HContact &=& \tL^{} d_1^\dag c^{}_{1,L} + \tL^* c_{1,L}^\dag d^{}_1  + \tR^{} d_{\MS/2+1}^\dag c^{}_{1,R} + \tR^* c_{1,R}^\dag d^{}_{\MS/2+1},
\end{eqnarray}
\end{widetext}
where $n_\ell = d_\ell^\dag d_\ell$ is the local density operator of site $\ell$, 
and $n_{k,\alpha}=c_{k,\alpha}^\dag c_{k,\alpha}$ is the density operator for momentum level $k$ in lead $\alpha$.
$\MS=6$ denotes the number of sites in the ring, $M$ the number of real-space sites in the leads, and $k$ labels 
the momentum-space sites of the leads. In this work we use the values $t_i=\tDot$ on the dot and give values 
of $\tBC$ in units of $\tDot$. Mostly we consider the case $\tBC=-\tDot$ corresponding to half a flux-quantum 
through the ring, $\Phi=\Phi_0$. Further, we use the values $\tL=\tR = t'=0.5\tDot$ for the coupling to the leads. 
Additionally we use a combination of a logarithmic discretization to cover a large energy-scale of the band, 
and switch to a linear discretization for the low-energy sector close to the Fermi edge \cite{Bohr_Schmitteckert:PRB2007},
for details of discretization issues we refer to \cite{Schmitteckert:JPCS2010}.
We model each lead by a real-space tight-binding chain coupled to a momentum-space part\cite{Bohr_Schmitteckert:PRB2007,Bohr:2007}, 
an advantageous setup for representing all relevant energy scales of a given problem. 
Within this setup we evaluate the Kubo formula for conductance, 
explicitly the two correlators\cite{Bohr_Schmitteckert_Woelfle:EPL2006}
\begin{eqnarray}
  g_{J_iJ_j} &=&
  \frac{e^2}{h}\left<\psi_0\right| J_i \frac
  {8\pi\eta(\HH-E_0)}{\big((\HH-E_0)^2+\eta^2\big)^2} J_j \left|\psi_0\right>,\label{JJ KuboCorrelator} \\
  g_{J_iN} &=& -\frac{e^2}{h} \left<\psi_0\right| J_i \frac{4\pi i\eta}{(\HH-E_0)^2+\eta^2} N
  \left|\psi_0\right>, \label{JN KuboCorrelator}
\end{eqnarray}
where $J_\ell=i(t_\ell^*c_{\ell-1}^\dag c_\ell - t_\ell c_\ell^\dag c_{\ell-1})$ 
denotes the current operator on site $\ell$ and $N=\frac 12 (N_L-N_R)$ is the rigid shift of the levels
in the leads corresponding to the applied voltage perturbation. 
Note that the Hamiltonian $\HH$ in Eqs.~(\ref{JJ KuboCorrelator},\ref{JN KuboCorrelator}) 
contains all interactions and couplings but not the voltage perturbation, 
and $|\psi_0\rangle$ is the ground state of this Hamiltonian. 

\subsection{Degenerate ground-state}
In the case where the ground-state $|\psi_0\rangle$ is (near-) degenerate the evaluation of the Kubo formula sketched above breaks down. 
Rather than a single ground-state correlator at zero temperature, a finite temperature multiple `ground-state' average must be used,
since in general it cannot be decided numerically whether a finite gap is a physical gap, or caused by numerical inaccuracies.
The evaluation of the Kubo-formula in this case is thus
\begin{eqnarray}
  g &=& \frac{1}{Z}\sum_{n=1}^N e^{-\beta(E_n-E_0)} g_n,
\end{eqnarray}
where $g_n$ is the conductance of the $n$'th (near-) degenerate ground-state level calculated 
using Eqs.~(\ref{JJ KuboCorrelator},\ref{JN KuboCorrelator}), $\beta$ is the inverse temperature,  
$E_n$ is the energy of the $n$'th level, and $Z = \sum_{n=1}^N e^{-\beta(E_n-E_0)} $ is the partition sum. 
Here we set $\beta$ of the order of the inverse level spacing of the leads, thus
averaging over low lying states having an excitation gap smaller than the finite size resolution of the leads. 
In addition $N$ is chosen sufficiently large to cover all relevant (near-) degenerate states.

\section{Results}
It was shown in \cite{Buettiker_Imry_Azbel:PRA1984,Gefen_Imry_Azbel:PRL1984} that a continuum model of an 1D ring of non-interacting spinless fermion
threated by a flux $\pi$ shows  antiresonances at flux $\phi=\pi$ leading to new phenomena in the high temperature limit when
interaction is turned on.\cite{Dmitriev_Gornyi_Kachorovskii_Polyakov:arxv2009}
In the non-interacting limit the transport properties of the benzene-like ring-structure sketched in Fig.~\ref{Ring-Lead-Setup} 
can be calculated exactly 
and in the infinite lead limit the conductance is identically zero for all gate voltages, for all couplings $t'$ and at all temperatures
due to a perfect interference between the two paths through the ring. 
Note that on a lattice this property does not hold for a ring consisting of four or eight sites.
Using DMRG to evaluate the Kubo formula for conductance we calculate the conductance for different values of the strength of the interaction. We typically use $400-800$ states per block in the DMRG procedure, and the momentum-space part of the lead is described by $40$ logarithmically discretized levels to cover the broad energy-range of the band, and additionally $10$ linearly discretized levels close to the Fermi-edge to ensure a good discretization here.

\begin{figure}[htb]
	\includegraphics[angle=-90,width=0.5\textwidth]{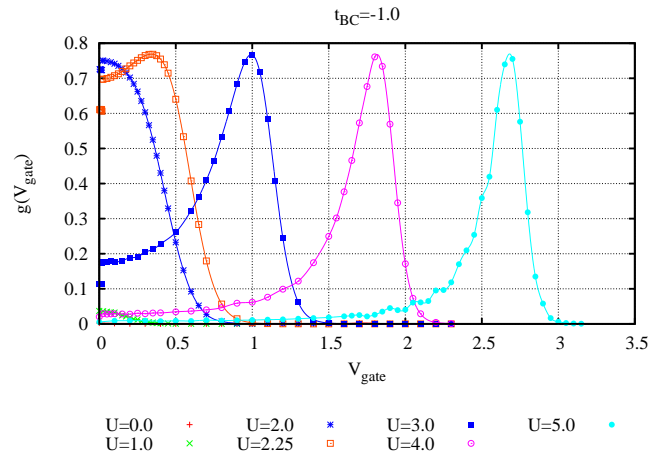}
	\caption{\ColorOnline Conductance versus gate-potential for the ring-structure varying the nearest-neighbor interaction strength $U$.
	In the non-interacting limit the conductance is zero to the precision of our finite-size setup, whereas the values for moderate
	interaction strengths approaches the unitary limit. 
	Lines are added to the DMRG data as guides to the eye.}\label{Conductance_vs_gate_Plot}
\end{figure}

The conductances obtained are shown in Fig.~\ref{Conductance_vs_gate_Plot}.
In the non-interacting limit we do indeed find a vanishing conductance with only minor finite-size deviations, 
originating from the finite size of the lead used in the DMRG setup. 

Increasing the strength of the interaction, $U$, the values of the conductance also increase,
and eventually a resonance is formed at zero gate-potential.
For the interaction strength $U=2$ the `resonant' value of the conductance is thus found to be $g\approx 0.75$. 
The shape of this resonance\, i.e.\ the exponential decay of the resonance with the gate voltage, differs significantly from the Lorentzian shape usually found 
in simple resonant systems\cite{Bohr_Schmitteckert_Woelfle:EPL2006,Bohr:2007}, 
indicating that a more complicated mechanism is at play.

Increasing the strength of the interaction further, $U=2.25$, a broad `plateau' in the conductance is formed around zero gate-potential. 
The plateau is significantly wider than a Lorentzian of the corresponding height, and resembles somewhat a split Kondo resonance, 
the splitting introduced by the hopping to the leads that also allows for transport. 
\begin{figure}[htb]
	\includegraphics[width=0.5\textwidth]{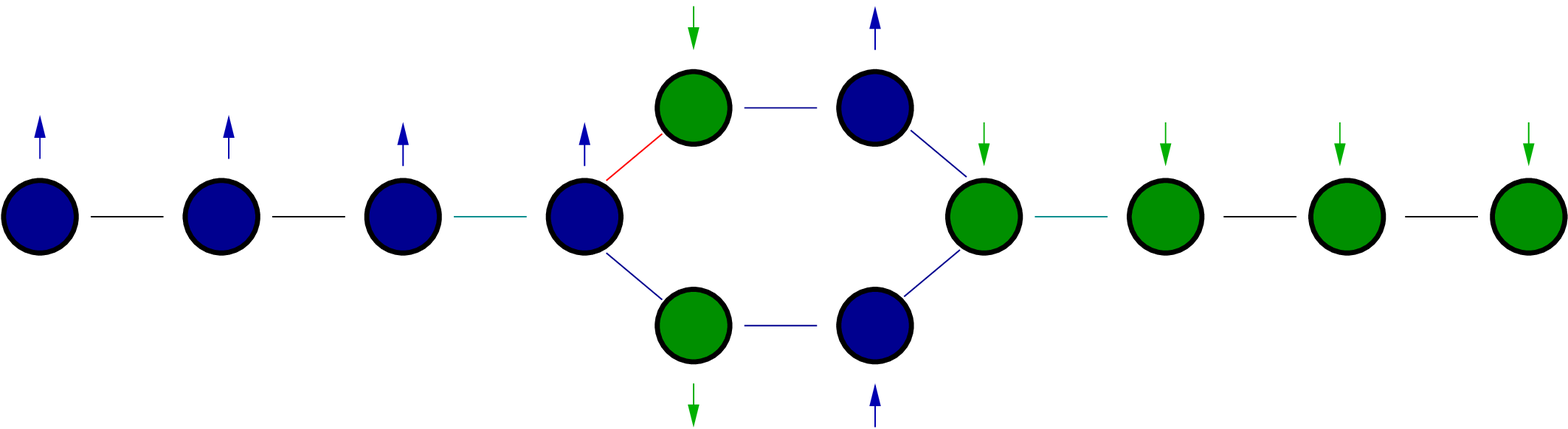}
	\caption{\ColorOnline Conjectured Kondo setup. }\label{KondoSketch}
\end{figure}
To show the similarity to the single impurity Anderson model (SIAM) we label in 
Fig.~\ref{KondoSketch} 
the left (right) lead as up (down) electrons. Setting $\tDot=0$ no mixing of `up' and `down' states occurs,
and  strong interaction forbids adding/removing an additional particle. 
In order to have transport it is necessary to switch on the hybridization
between the dot sites, which acts like a magnetic field suppressing the proposed geometric Kondo effect. 
In our case, due to the added flux,
the degeneracy of the single-particle levels is not lifted leaving room for Kondo physics. 
Nevertheless, the hybridization of the dot provides a mixing of the up and down states proposed in  Fig.~\ref{KondoSketch}, 
which is necessary to enable transport through the ring. 
However, by increasing $U$ a charge density wave (CDW)
ordering again becomes preferred when the interaction reestablishes two well separated states. It is interesting to note, that the effect is
most dominant for interaction values close to where a phase transition to a CDW ordered state appears in the thermodynamic limit
at $U_{\mathrm c}=2 t$. 

\begin{figure}[hbtp]
	\includegraphics[angle=0,width=0.5\textwidth]{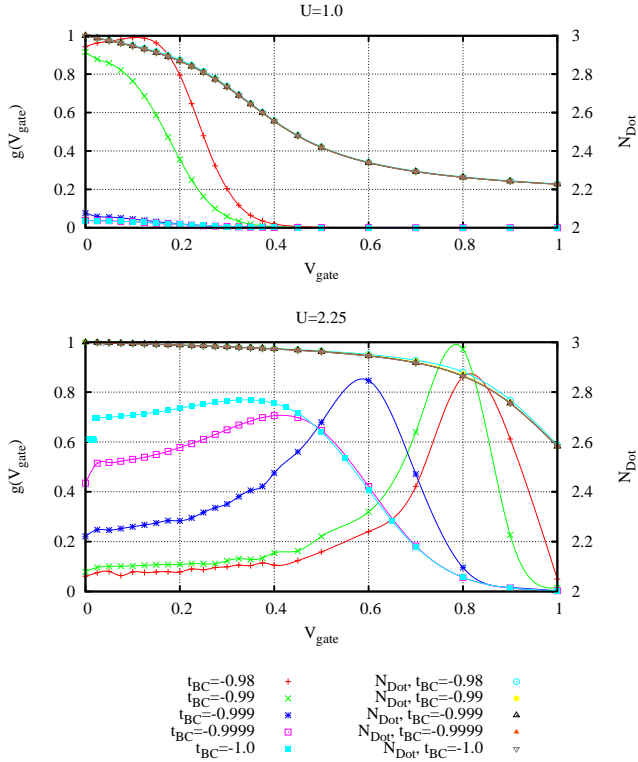}
	\caption{\ColorOnline An asymmetry of the hopping in the ring rapidly destroys the geometric Kondo-effect. 
	This figure shows the effect on the conductance and the occupation of the ring when modifying a single hopping by 0.01-2\%. 
	When the geometric Kondo-effect is destroyed a normal resonant structure pattern is rediscovered, 
	and the resonance moves towards the normal position where the occupation is half-integer. 
	Lines are added to the DMRG data as guides to the eye.}\label{Asymmetric_convergence}
\end{figure}

In order to test this idea we have performed similar calculations on slightly asymmetric rings, 
reducing the magnitude of a single hopping slightly from unity, breaking the degeneracy between the single particle levels, 
and obtained the conductances shown in Fig.~\ref{Asymmetric_convergence}.
As the figure shows for the interaction strength $U=2.25$, the introduced asymmetry rapidly destroys the effect, as expected for Kondo physics.
Ignoring the terms mixing the two levels at the Fermi surface with the other four levels we can map the six site
ring on a single Impurity Anderson model with an effective interaction of $\tilde{U}=2U/3$ and an effective hybridization
of $\tilde{t}' = t'/\sqrt{6}$ as displayed in Fig.~\ref{KondoSketch}.
The Kondo temperature \TK, at which the Kondo resonance of the SIAM is destroyed is given by 
$\TK=D\exp( - 1/J(U)  )$, see \cite{Hewson:1997}, where $D=2t$ is the band cut-off, and $J={t'}^2 \left( 1/|\epsilon_d|  + 1/|\epsilon_d + U|\right)$ 
is the effective Kondo coupling. 
Setting $\tilde{t}'=0.5t / \sqrt{6}$ and $\epsilon_d=-\tilde{U}/2$ due to the particle-hole symmetry,
we obtain for $U=2.25t$ a Kondo temperature of $\TK \approx 2.5\cdot10^{-4}\,t$, 
which is in reasonable agreement with our numerical results.

Also plotted in Fig.~\ref{Asymmetric_convergence} is the total density of the ring for the various parameter choices.
Remarkably the electron-density of the ring remains virtually unchanged when the asymmetry is varied, 
although the conductance of the ring changes significantly. This clearly demonstrates that the observed effect 
is an interference-effect. Furthermore, increasing the asymmetry of the ring and thereby destroying the interference effect, 
the resonance is pushed towards the usual location for resonant structures, where the particle number on the structure is half-integer, 
and at the same time the line-shape becomes increasingly Lorentzian, although an asymmetry with a long tail persists.

In the case $U=1.0$, where no geometric Kondo-effect is present, reducing the symmetry in the ring also results in changes.
From the symmetric case, $\tBC=-\tDot$, where only a very small resonance is found, to the most asymmetric case considered here, 
$\tBC=-0.98\tDot$, a clear resonance develops. 

\begin{figure}[htb]
	\includegraphics[width=0.5\textwidth]{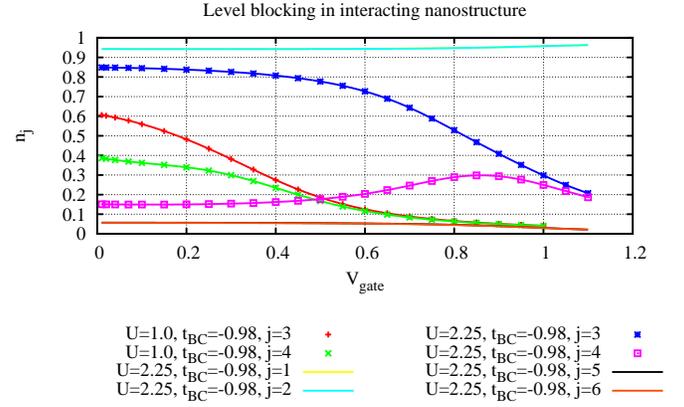}
	\caption{\ColorOnline Occupation $n_j = \langle { \tilde{c}^+_j \tilde{c}^{}_j } \rangle$ 
	of the $j$-th single-particle levels of the non-interacting, isolated 
      ring induced by the coupling to the leads and the interaction on the ring. }\label{SingleLevelOccupation}
\end{figure}

We explain this new resonance as the result of a level blocking mechanism, similar to \cite{Goldstein_Berkovits_Gefen_Weidenmueller:2008}:
When applying a gate voltage the upper level of the two levels close to the Fermi surface of the leads is  pushed out of resonance first.
The second level remains occupied since emptying the level would cost interaction energy due to the particle-hole symmetric interaction. 
Therefore adding or removing a particle costs interaction energy and the lower level remains occupied although it is  pushed above the Fermi level. 
This is observed in Fig.~\ref{SingleLevelOccupation}, where the single-particle occupations are plotted for the two cases considered. For $U=1.0$ the level $j_3$ remains occupied, whereas the level $j_4$ is emptied much faster.
Since the levels now have different occupations the perfect destructive interference is destroyed and a conductance peak appears at the position,
where the lower level is on resonance. 
By further increasing the gate voltage we finally empty the lower level as well.
Interestingly, for $U=2.25$ the filling of the upper level increases during this intermediate regime. 
Finally, for large $V_{\rm gate}$ both levels share the same occupation again and destructive interference is reestablished. 

A note is in order about the calculations for small $V_{\mathrm{gate}}$: In a region around $V_{\mathrm{gate}} = 0$ 
the combined lead and ring system is effectively degenerate, and thus the degenerate method is applied in this parameter range. 
However, even using this expression the evaluation of the conductance for small gate-potentials remains difficult due to numerically difficult resolvent equations, 
and the results obtained there are not expected to be accurate. 

\section{Discussion}
At the fundamental level the results shown in this work clearly demonstrate that the simple picture of individual electrons passing through the transport region one after the other gives a significantly different result 
than when including even moderate electron-electron interactions. Rather a complicated many-body interference effect is formed, and we have proposed/conjectured a Kondo-effect as the explanation for the remarkable line-shape observed. 
The proposed Kondo-effect lies in the geometrical degree of freedom, and hence differs from the standard Kondo-effect in the spin-degree of freedom. Introducing an asymmetry in the ring clearly destroys the effect, in a manner similar 
to the effect of a magnetic field on the standard Kondo-effect. Evidence for the proposed Kondo effect is given by the disappearance of the conductance peak through
a weak effective magnetic field, while the peak is robust against a small gate voltage. 
While at first sight our model seems to be impractical for experimental realizations as one cannot thread a benzene ring with half a flux quanta, such a model can actually
appear as effective models in molecular electronics \cite{Sense:X2011}.

\section*{Appendix} \label{sec:U0}
The solution of the non-interacting system can most easily be calculated by scattering theory.
There one searches for scattering states of the form $\e^{\im k x} + r \e^{-\im k x}$ in the left lead, 
and $\tau \e^{\im kx}$ in the right lead, and $\Psi_n$, $n=0,1,\ldots,5$ for the wave function on the dot,
which fulfill the Schr{\"o}dinger equation $(\HH - E) \Psi = 0$. In the leads the solution of the tight binding
leads requires $E(k) = -2 t \cos(k)$. Solving the linear set of equations for flux $\phi=\pi$ and an incoming wave 
at the Fermi surface of $k=\pi/2$ immediately leads to $\tau=0$ for all gate voltage $\Vg$.
Therefore the linear conductance $G = | \tau |^2 = 0$ vanishes for arbitrary $\tL$ and $\tR$.
In contrast, setting the flux to zero one obtains a finite conductance $G= \frac{ 16 t^4 \tL^2 \tR^2}{ \left(4 t^4 + \tL^2 \tR^2 \right)^2 } \frac{e^2}{h}$
for the linear conductance at zero gate voltage.

In order to estimate the Kondo temperature in the strongly interacting case we define
\begin{eqnarray}
    \hat{N}_\uparrow   &=& \hat{n}_0 + \hat{n}_2 + \hat{n}_4 \\
    \hat{N}_\downarrow &=& \hat{n}_1 + \hat{n}_3 + \hat{n}_5 \,.
\end{eqnarray}
If we then  then look at the interaction term generated by
\begin{equation}
  \hat{U}_{\mathrm{alt.}} = -\frac{ \widetilde U}{2} \left( \hat{N}_\uparrow  - \hat{N}_\downarrow \right)^2 =  {\widetilde U} \hat{N}_\uparrow \hat{N}_\downarrow +  \frac{\widetilde U}{2}\left(\hat{N}_\uparrow  - \hat{N}_\downarrow \right) 
\end{equation}
we find that it matches our original interaction $\hat{U}$ up to the addition of terms with distance d=3, i.e.\ 0--3, 1--4, 2--5,
at least as long as we are close to the half filled dot.
In the case of a charge density wave like ordering these terms leads to the same contribution as the nearest neighbour interaction
and therefore have to reduce ${\widetilde U}$ to $2U/3$ in order to describe our system.

\ 
\acknowledgments

We would like to thank Ferdinand Evers 
for insightful discussions.
The DMRG calculations were performed on the HP XC4000 at the Steinbuch Center for Computing (SCC) Karlsruhe under project RT-DMRG, 
with support through the priority programme SPP 1243 of the DFG.
The work was performed while DB being at the Department of Physics and Astronomy of the University of Basel.

\bibliography{Benzene} 

\end{document}